\documentclass{iau} 
\usepackage{graphicx}

\title[Age-metallicity structure of the Milky Way disc] 
{The age-metallicity structure of the Milky Way disc with APOGEE}

\author[J. Ted Mackereth et al.]   
{J. Ted Mackereth$^1$,
 Jo Bovy$^2$ and Ricardo P. Schiavon$^1$ and the SDSS-IV/APOGEE Collaboration}

\affiliation{$^1$Astrophysics Research Institute, Liverpool John Moores University, \\ 146 Brownlow Hill, Liverpool, L3 5RF, United Kingdom \\ email: {\tt J.E.Mackereth@2011.ljmu.ac.uk} \\[\affilskip]
$^2$Astronomy and Astrophysics Department, University of Toronto, \\ 50 George Street, Toronto, ON M5S 3H4, Canada \\email: {\tt bovy@astro.utoronto.ca}}

\pubyear{2017}
\volume{334}  
\jname{Rediscovering our Galaxy}
\editors{C. Chiappini, I. Minchev, E. Starkenburg \& M. Valentini, eds.}
\begin{document}

\maketitle

\begin{abstract}
The best way to trace back the history of star formation and mass assembly of the Milky Way disc is by combining chemical compositions, ages and phase-space information for a large number of disc stars. With the advent of large surveys of the stellar populations of the Galaxy, such data have become available and can be used to pose constraints on sophisticated models of galaxy formation. We use SDSS-III/APOGEE data to derive the first detailed 3D map of stellar density in the Galactic disc as a function of age, [Fe/H] and $\mathrm{[\alpha/Fe]}$. We discuss the implications of our results for the formation and evolution of the disc, presenting new constraints on the disc structural parameters, stellar radial migration and its connection with disc flaring. We also discuss how our results constrain the inside out formation of the disc, and determine the surface-mass density contributions at the solar radius for mono-age, mono-[Fe/H] populations. 

\end{abstract}

\firstsection 
\section{Introduction}

The field of Galactic Archaeology is underpinned by the necessity for robust measurements of the present day structure of the Galaxy, which provide strong constraints on any model for its formation and evolution. The disc of the Milky Way is among the more difficult to observe of its components, due to the dust extinction along lines of sight from our position inside it. Spectroscopic surveys in the near infra-red (NIR) such as the Apache Point Observatory Galactic Evolution Experiment (\cite[APOGEE, Majewski et al. 2015]{majewskietal2017}), cut through this dust, and are measuring the chemistry of large numbers of its stars with very high precision, allowing the mapping of the stellar chemical abundances (e.g. \cite[Nidever et al. 2015]{nidever2015}, \cite[Hayden et al. 2015]{hayden2015}) and density structure of the disc (\cite[e.g. Bovy et al. 2016b]{bovy2016b}) across many Galactocentric radii.

\section{Data \& Method}

In this work, we measure the disc structure as a function of age and metallicity, applying an adaptation of the density fitting methodology developed in \cite[Bovy et al. 2016a]{bovy2016a},\cite[b]{bovy2016b}, to the APOGEE DR12 data \cite[(Holtzman et al. 2015)]{holtzman2015}. Ages are measured for a large sample of APOGEE DR12  stars via modelling of the $\mathrm{[C/N]}$ - mass relation (\cite[Martig et al. 2016]{martig2016}). Distances, computed using a Bayesian estimation based method, are taken from \cite[Hayden et al. (2015)]{hayden2015}. We use the full red giant branch (RGB) sample, after correcting for the selection function of these stars in APOGEE  \cite[(see Zasowski et al. 2013, for information regarding targeting in APOGEE)]{zasowski2013}. We divide stars in the $\mathrm{[\alpha/Fe]}$-$\mathrm{[Fe/H]}$ plane into a high- and low-$\mathrm{[\alpha/Fe]}$ population by visually identifying the boundary between the two populations. Stars are then binned by age and metallicity assuming $\Delta \mathrm{age} = 2$ Gyr, and $\Delta \mathrm{[Fe/H]} = 0.1$ dex. The density structure of the populations in these bins is fit assuming a broken exponential radial surface density profile with an inner and outer scale length $h_{R,\mathrm{[in/out]}}$ on either side of a peak radius $R_\mathrm{peak}$. The vertical profile is fit by a single exponential with scale height at the solar radius $h_Z$. We also invoke a `flaring' term, such that the scale height of the vertical profile changes exponentially with radius, with a scale length of $R_{\mathrm{flare}}$. The assumption of a broken exponential radial profile does not preclude a fit by a single exponential, if the data are better fit in this way. The full methodology and results are described in detail in \cite[Mackereth et al. (2017)]{mackereth2017}. Tests of the method are also performed, accounting for the blurring of results by the age errors, which is found not to affect our conclusions.

\section{Results}
\begin{figure}
\includegraphics[width=\textwidth]{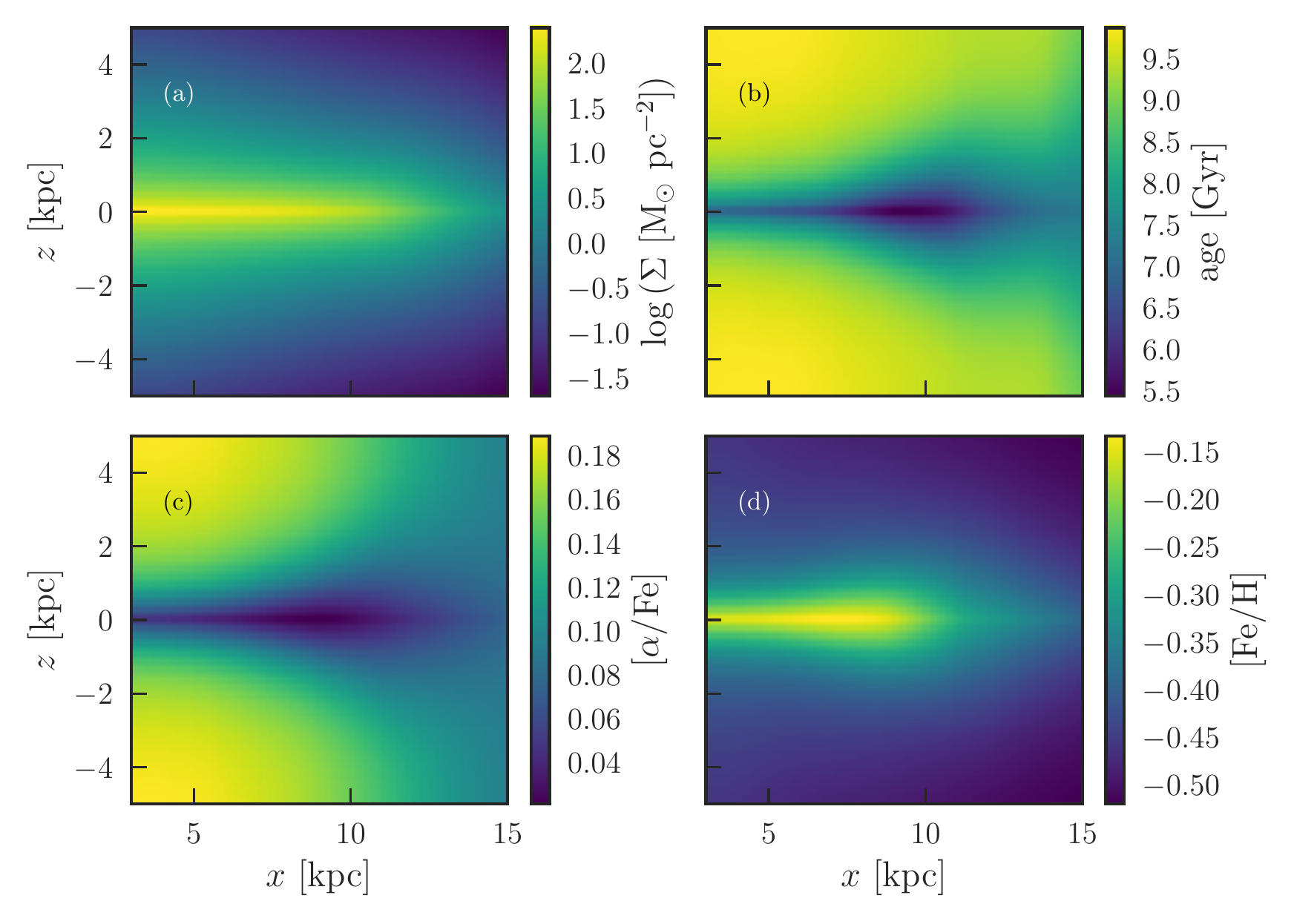}
\caption{\label{fig:maps} Edge on views of the Milky Way disc, corrected for survey selection effects, from combining the parametric density models for mono-age mono-$\mathrm{[Fe/H]}$ populations, weighted by their contribution to the surface-mass density at the solar radius (assumed to be $8$ kpc in this work). Panel (a) shows the surface-mass density map, panel (b) shows the surface-mass density weighted mean age map, (c) shows mean $\mathrm{[\alpha/Fe]}$ and (d) shows mean $\mathrm{[Fe/H]}$, all as a function of galactocentric $x$ and $z$. The flared, young, low $\mathrm{[\alpha/Fe]}$ disc is visible in the age and  $\mathrm{[\alpha/Fe]}$ maps, but no discernible flared disc feature is visble in the total surface-mass density map, which looks qualitatively like a thin plus thick disk.}
\end{figure}

Fitting each of the mono-age mono-$\mathrm{[Fe/H]}$ populations allows us to study how the structure of the disc changes as a function of age and metallicity, yielding insights into the history of the disc, and on models of disc formation. 

{\underline{\it Radial surface density profiles}}. We find that high $\mathrm{[\alpha/Fe]}$ populations are best fit radially as a single exponential, with a scale length of $\sim 1.9$ kpc which is fairly insensitive to age and metallicity, and in good agreement with other studies of the high $\mathrm{[\alpha/Fe]}$ disc (e.g., \cite[Cheng et al. 2012]{cheng2012}). The low $\mathrm{[\alpha/Fe]}$ populations are all fit by a broken exponential whose peak radius, $R_{\mathrm{peak}}$, and inner and outer scale lengths, $h_{R,\mathrm{[in/out]}}$, are a function of both age and metallicity. The width of the profiles around the peak broaden significantly with increasing age of the population, which we interpret as a signature of radial migration or radial heating of the disk. $R_{\mathrm{peak}}$ increases with decreasing $\mathrm{[Fe/H]}$, which is indicative of the radial metallicity gradient in the disc. We find that the slope of the relation declines with increasing age, indicating the shallowing of the metallicity gradient with age, as found also by, e.g., \cite[Anders et al. (2017)]{anders2017}.

{\underline{\it Vertical profiles}}. We find that the vertical component of each mono-age mono-[Fe/H] population is best fit by a single exponential at each Galactocentric radius. Both high- and low-$\mathrm{[\alpha/Fe]}$ mono-age populations show evidence for flaring, which is stronger in low-$\mathrm{[\alpha/Fe]}$ populations. We show that the surface-mass density weighted-distribution of $h_Z$ is a declining exponential function of $h_Z$, confirming the findings of \cite[Bovy et al. 2012]{bovy2012} that the vertical structure of the disc is smooth and continuous, with no signal of a distinct thick disc.

{\underline{\it Edge on maps of the disc}}. In Fig. \ref{fig:maps} we combine all the fitted density models for high and low $\mathrm{[\alpha/Fe]}$ and for all ages and metallicities. We combine different populations by correcting the observed star counts of giants for the selection function of APOGEE (see above) and by extrapolating the number of giants observed by APOGEE to the full present-day-mass range of the stellar populations sampled by the different bins. The latter is done using the PARSEC stellar evolution models (\cite[Bressan et al. 2012]{bressan2012}). This procedure provides the full stellar mass density profile for each bin and summing over the bins provides the total stellar density profile of the stellar disk. Integrating this combined cylindrically symmetric 3D model in the $y$ axis gives a 2D map in the $x-z$ plane, from which the mean age, $\mathrm{[\alpha/Fe]}$ and $\mathrm{[Fe/H]}$ in each pixel can be recovered. This provides a selection-function-corrected edge-on view of the Milky Way, in the absence of extinction. The flaring of young, low-$\mathrm{[\alpha/Fe]}$ populations, which creates a negative radial age gradient above the midplane, stands out in that edge-on view of the Galactic disc. The total surface-mass density map has a structure which is qualitatively consistent with those of external galaxy thick discs seen edge-on (\cite[e.g., Yoachim \& Dalcanton 2006]{yoachim2006}).

\section{Implications and Future Work}

The precise measurement of the Galactic disc structure as a function of age and metallicity is strongly constraining to models for the formation and evolution of the Galaxy. As an example, any proposed model for the emergence of discontinuities in the $\mathrm{[\alpha/Fe]}$-$\mathrm{[Fe/H]}$ plane -- namely, the bimodality in the $\mathrm{[\alpha/Fe]}$ distribution at fixed $\mathrm{[Fe/H]}$, as observed by various studies in the solar vicinity (e.g. \cite[Fuhrmann et al. 1998]{fuhrmann1998}, \cite[Bensby et al. 2003]{bensby2003},  \cite[Prochaska et al. 2000]{prochaska2000}, \cite[Bensby et al. 2005]{bensby2005}, \cite[Adibekyan et al. 2012]{adibekyan2012}, \cite[Bensby et al. 2014]{bensby2014}) and beyond (e.g \cite[Anders et al. 2014]{anders2014}, \cite[Nidever et al. 2015]{nidever2015}, \cite[Hayden et al. 2015]{hayden2015}) -- must produce a disc whose vertical structure is continuous at the solar radius, but whose radial structure varies with $\mathrm{[\alpha/Fe]}$. There are ongoing theoretical attempts to reproduce features of the disk in terms of spatial and elemental abundance structure (e.g.  \cite[Brook et al. 2004]{brook2004}, \cite[Chiappini et al. 2009]{chiappini2009}, \cite[Brook et al. 2012]{brook2012}, \cite[Minchev et al. 2013]{minchev2013}, \cite[Martig et al. 2014a]{martig2014a}\cite[,b]{martig2014b}, \cite[Andrews et al. 2015]{andrews2015}, \cite[Toyouchi \& Chiba 2016]{toyouchi2016}, \cite[Ma et al. 2017]{ma2017}), and these will be an important component in developing our understanding of these results. 

Future releases of data from APOGEE and other surveys will further constrain the structure of the disc as a function of chemical composition and age, both by increasing the number of stars observed and through refinement of techniques for measuring stellar abundances, and in particular, ages. The determination of precise ages with uncertainties as low as $3\%$ (as expected in the proposed PLATO mission, \cite[Miglio et al. 2017]{miglio2017}), will allow a deeper understanding of the older disc components, shedding light on an important period in its evolutionary history. In addition to this, the ESA-\emph{Gaia} mission (\cite[Gaia Collaboration et al. 2016a]{gaia2016a}), whose first data were released last year (\cite[Gaia Collaboration et al. 2016b]{gaia2016b}), will provide detailed 6D phase-space information for many of the stars in these surveys, which will allow for the addition of more strict kinematic and dynamic constraints on these models.

\begin{acknowledgements}
\begin{scriptsize}
Funding for SDSS-III was provided by the Alfred P. Sloan Foundation, the Participating Institutions, the National Science Foundation, and the U.S. Department of Energy Office of Science. The SDSS-III web site is http://www.sdss3.org/.
\end{scriptsize}
\end{acknowledgements}

%
%
%
%
%
%

\end{document}